\documentclass[12pt,preprint]{aastex}

\usepackage[percent]{overpic}
\usepackage{graphicx}
\usepackage{natbib}

\newcommand{\kms}{\hbox{km~s$^{-1}$}}
\newcommand{\kmsmpc}{\hbox{km~s$^{-1}$~Mpc$^{-1}$}}

\begin{document}

\title{Dual black hole associated with obscured and unobscured AGN: CXO J101527.2+625911}

\author{
D.-C. Kim\altaffilmark{1}, 
E. Momjian\altaffilmark{2},
Ilsang Yoon\altaffilmark{1},
Minjin Kim\altaffilmark{3, 4},
A. S. Evans\altaffilmark{1, 5},
Ji Hoon Kim\altaffilmark{6, 7},
S. T. Linden\altaffilmark{5},
L. Barcos-Munoz\altaffilmark{1}, \&
G. C. Privon\altaffilmark{8}
}
\altaffiltext{1}{National Radio Astronomy Observatory, 520 Edgemont Road, Charlottesville, VA 22903, USA: dkim@nrao.edu, iyoon@nrao.edu, aevans@nrao.edu, stl7ey@virginia.edu, lbarcos@nrao.edu}
\altaffiltext{2}{National Radio Astronomy Observatory, P.O. Box O, Socorro, NM 87801, USA: emomjian@nrao.edu}
\altaffiltext{3}{Korea Astronomy and Space Science Institute, Daejeon 305-348, Korea}
\altaffiltext{4}{Department of Astronomy and Atmospheric Sciences, Kyungpook National University, Daegu 702-701, Korea: mkim.astro@gmail.com}
\altaffiltext{5}{Department of Astronomy, 530 McCormick Rd., University of Virginia, Charlottesville, VA 22904}
\altaffiltext{6}{Metaspace, 36 Nonhyeon-ro, Gangnam-gu, Seoul 06321, Korea: jhkim@naoj.org}
\altaffiltext{7}{Subaru Telescope, National Astronomical Observatory of Japan, 650 North Aohoku Place, Hilo, HI 96720, USA}
\altaffiltext{8}{Department of Astronomy, University of Florida, 211 Bryant Space Sciences Center, Gainesville, FL 32611, USA: george.privon@ufl.edu}

{\centering
  \textbf{Abstract}\par
}
We report the results of an investigation to determine the nature of the offset active galactic nucleus (AGN) found in the source CXO J101527.2+625911. 
Hubble Space Telescope and Chandra X-ray observatory data had suggested that the offset AGN, which has an angular separation of only 0\farcs26 from the center of the host galaxy, is a recoiled Super Massive Black Hole (rSMBH). 
We carried out high angular resolution observations with both the VLBA (1.54 GHz) and the VLA (10.0 GHz \& 33.0 GHz) and detected a single compact radio source in the center of the host galaxy, 
with no radio continuum emission associated with the offset AGN. 
The detected radio source has a high brightness temperature value of $T_b=7.2\times10^{7}$ K, indicating that the radio emission is associated with an AGN. 
Furthermore, we present the decomposition of high-resolution KECK spectra of the  [O III]5007\AA line into two narrow emission line components, 
which is a characteristic sign of a dual black hole system. 
These new radio and optical wavelength results suggest that CXO J101527.2+625911 is the host of a dual black hole system rather than a rSMBH.



\section{Introduction}

A dual/binary black hole system can be formed from two interacting galaxies.
The fate of the two black holes depends on a number of factors.
They could merge together to form a single supermassive black hole (SMBH), or
could remain a dual black hole system if the secondary galaxy looses significant fraction of its mass due to the tidal stripping effect of the primary galaxy at kpc scale (e.g. Callegari et al. 2009).
Recent simulations predict that the merged SMBH can attain a recoil velocity of 
a few hundred to a few thousands km s$^{-1}$ depending on mass ratios, spin magnitudes, and spin orientations of the merging SMBHs (Campanelli et al. 2007; Schnittman 2007; Baker et al. 2008; Lousto \& Zlochower 2011).
If the merging SMBHs are of equal mass and highly-spinning, and their spins are aligned along the orbital plane (superkick configuration), 
recoil velocities as large as 4000 km s$^{-1}$ (Campanelli et al. 2007) to 5000 km s$^{-1}$ (Lousto \& Zlochower 2011) can be reached
and the recoiling SMBH (hereafter rSMBH) will eventually escape from the host galaxy (i.e., Merritt et al. 2004).
Studying the rSMBH is important since they will provide an essential information about how the mergers grow 
their bulge and black hole mass to evolve toward the normal ellipticals.  


From imaging and spectroscopic surveys of the rSMBHs, 
'CXO J101527.2+625911' was identified as a potential rSMBH candidate (Kim et al. 2017).
CXO J101527.2+625911 is a QSO hosted merger remnant located at $z$=0.3504.
Its morphological type is a bulge-dominated elliptical, with two nuclei in its center - a bright
northern nucleus harboring AGN and is spatially offset by 1.26 kpc (0.256\arcsec) from the less bright southern nucleus that was identified as the center of the host galaxy. Spectral decomposition of the
KECK spectra suggests H$\beta$ broad emission line is redshifted by 175$\pm$25 km s$^{\rm -1}$ with respect to the systemic velocity (Kim et al. 2017).

The observed spatial and velocity offsets suggest that this galaxy could be hosting a rSMBH. However, there is a possibility that this system could be a dual/binary SMBH where the second black hole is hidden behind the dust in the center of the host galaxy. To resolve whether this system has a rSMBH or a dual/binary black hole system, we have carried out Very Long Baseline Array (VLBA) and Karl G. Jansky Very Large Array (VLA) radio continuum observations.

In this paper, we report the results of our VLBA and VLA observations.
The observations and data reduction are described in Section 2, 
the discussion is presented in Section 3, and a summary is presented in Section 4. 
Throughout this paper, the cosmology H$_0$ = 70 \kmsmpc, $\Omega_M$ = 0.3, and $\Omega_\Lambda$ = 0.7 is adopted.
At the distance of the target source, 1\arcsec\ subtends 4.94 kpc on the sky.
 
\section{Observations and Data Reduction}

The observations of the target source CXO J101527.2+625911 were conducted with the
VLA and the VLBA of the National Radio Astronomy Observatory (NRAO)\footnote{The National Radio Astronomy Observatory is a facility
of the National Science Foundation operated under cooperative agreement by Associated Universities, Inc.}. 
Table 1 summarizes the observations. Details of the observations with both facilities are described separately below.


\subsection{VLBA}
The VLBA observations of CXO J101527.2+625911 were carried out on 2017 October 15. We utilized the ROACH Digital Backend
and the polyphase filterbank (PFB) digital signal-processing algorithm to deliver eight 32 MHz data channel pairs, both with
right- and left-hand circular polarizations. The data recorded at each station were sampled at 2 bits. 
The total observing time was 3.56 hours, and the total bandwidth was 256 MHz centered at 1.54 GHz.

The observations utilized nodding-style phase referencing with a cycle time of 3 minutes: 
two minutes on the target source and one minute on the phase calibrator J1019+6320. 
This calibrator is located at an angular distance of $0.6^{\circ}$ from the target source CXO J101527.2+625911. 
The strong calibrator source 3C 147 was observed as a fringe finder and to calibrate the bandpass response.
The amplitude scale of the data was calibrated using measurements of the antenna gain and the system temperatures.
The data were correlated with the VLBA software correlator (Deller et al. 2011) in Socorro, New Mexico, with a 4 second correlator integration time.

In phase referenced observations, such as those presented here, the accuracy in the position of the phase calibrator determines the
accuracy of the absolute position of the target source (Walker 1999). 
The uncertainty in the position of the phase calibrator J1019+6320
is 0.36 mas and 0.72 mas in right ascension and declination, respectively (Ma et al. 2009). 
Furthermore, phase referencing, as employed in our observations, is known to preserve the absolute astrometric positions to better than ±10 mas (Fomalont 1999).

The editing, calibration, and imaging of the data were performed using the Astronomical Image Processing System (AIPS; Greisen 2003)
following standard Very Long Baseline Interferometry data reduction procedures. 
The final VLBA continuum image was made with a robust factor of 1 in the AIPS task IMAGR.

\subsection{VLA}
The VLA A-configuration ($B_{max}=36.4$ km) observations of CXO J101527.2+625911 were carried out on 2018 March 7 in the X (8 -- 12 GHz) and Ka (26.5 -- 40 GHz) frequency bands. 
Utilizing the 3-bit samplers, the observations delivered a total of 8 GHz bandwidth at Ka-band centered at 33 GHz, 
and a total of 4 GHz bandwidth at X-band centered at 10 GHz. 
The flux density scale calibrator and bandpass calibrator was 3C 286, and the complex gain calibrator was J0921+6215. 
Both X- and Ka-band data sets were edited and calibrated using the Common Astronomy Software Applications (CASA) package version 5.1. 
In order to improve the calibration for the science target, we also performed two rounds of phase-only self-calibration. 
The final continuum images were made with Briggs weighting and a robust factor of 0.5 in the CASA task tclean.

%

%

\section{Result and Discussion}

Fig. 1a shows Hubble Space Telescope (HST) Advanced Camera for Surveys (ACS) I-band (F775W) image of the nuclear region in CXO J101527.2+625911, 
where green plus marks are the positions of the two nuclei: a northern nucleus (offset AGN) and a southern nucleus (center of the host galaxy).
The white dot represents the center of the Chandra X-ray emission and the circle represents its positional uncertainty.
Fig. 1b is a zoomed-in image of Fig. 1a, where white and blue contours represent the continuum emission at 33 GHz measured with the VLA, and the 1.54 GHz measured with the VLBA.
Fig. 1c is a zoomed-in image of Fig. 1b, where synthesized beams are plotted on the bottom left (VLA) and bottom right (VLBA).
The synthesized beam sizes are 0\farcs0125$\times$0\farcs005 and 0\farcs06$\times$0\farcs05 for the VLBA and VLA 33 GHz images, respectively.
In both the VLBA and VLA observations, we have detected a single compact radio source near the center of the southern nucleus (host galaxy).
The coordinates and flux densities of the radio source are summarized in Table. 1.
Assuming the positional uncertainty of the VLA image is about 10\% of the FWHM of the synthesized beam, 
then the VLBA and VLA radio emissions are originating from the same location.
The absolute astrometry of the HST ACS images ($\sim$0\farcs3) is slightly larger than the nuclear separation (0\farcs265).
To identify where the location of the radio emission corresponds in the HST ACS image,
we first referenced the GAIA Data Release 2 Catalogue (Gaia Collaboration 2018).
The coordinates of the offset AGN (northern nucleus) identified from the GAIA Catalog are R.A.=10h15m27.2652s$\pm$0.0002s and Dec.=62d59$'$11.6095$''$$\pm$0.0003$''$.
Next, we registered this position with the northern nucleus in the HST image and measured the coordinates of the southern nucleus
and found that the coordinates are R.A.=10h15m27.264s$\pm$0.0002s and Dec.=62d59$'$11.345$''$$\pm$0.0003$''$.
Thus, the positional offset between the VLBA center and the center of the host galaxy is
about 0\farcs045 and corresponds to physical scale of 350 pc.

\subsection{A rSMBH or a dual/binary SMBH?}

Previously in Kim et al. (2017), we have identified this source as a possible recoiling supermassive black hole (rSMBH)
mainly because the X-ray emission is associated with the offset AGN and no sign of AGN emission at optical wavelengths is detected in the nucleus of the host galaxy, though a deeply buried AGN is still possible.
As discussed in the previous section, we have detected the radio emission near the center of the host galaxy.
In general, radio continuum emission may originate from starburts activities (SNe, outflows, shocks, etc...), or AGN activity.
The 1.54 GHz VLBA observations reveal a continuum source with a brightness temperature $T_b=7.2\times10^{7}$ K. 
Such a high brightness temperature value clearly
indicates that the radio emission is AGN related (e.g., Condon et al. 1991; Condon 1992).

The detection of an optically hidden AGN in the host galaxy, as revealed through the VLBA observations, suggests that this is a dual or binary black hole system
with an obscured AGN (southern nucleus) and an unobscured AGN (northern offset AGN).
A commonly adopted method for selecting black hole pair candidates is to identify a pair of narrow emission lines (e.g. [O III]5007 line) in the QSO spectra.
Thus, if this system is indeed a dual/binary black hole system, we could see two sets of narrow emission lines in the spectra.
In the SDSS spectra (Fig. 2a), it is hard to tell if there exist two kinematically separated narrow emission lines, since a single Gaussian component (blue line) fits well to the [O III]5007 line.
However, if we fit a single Gaussian to the [O III]5007 line in high-resolution KECK spectra (Fig. 2b),
we see an extra component in the red part of the spectra (indicated by the arrow in the plot) 
- this is possible evidence that two narrow emission lines could be closely overlapped. 
To test this, we tried spectral decompositions using one [O III] component line versus two [O III] component lines
and compared the results.
In the fit, we added an [O III] blue asymmetry component that is often observed in QSOs and interpreted as evidence of an outflow.
The fitting results are plotted in Figure 3a (one narrow component fit) and Figure 3b (two narrow components fit), 
where the black, cyan, blue, green, orange, and red lines in the plot represent
the observational data and fit to the power-law continuum, broad emission line, narrow emission lines, 
[O III] blue asymmetry component, and the combined model, respectively.
The dotted line represents an absorption line component.
It is found that the spectral decomposition with two narrow line components fit better ($\chi^2$=2.0) than that with a single narrow line component ($\chi^2$=6.7).
Our result emphasizes that the high-resolution spectral data is essential to identify closely-blended two narrow emission lines in dual black hole systems.

The blueshifted [O III] line could be coming from the offset AGN since
the line strength of the blueshifted line is stronger than that of the redshifted one.
The southern nucleus AGN is Compton thin (Juneau et al. 2011; log (L$_{\rm{X-ray}}$/L$_{\rm{[O III]}}$) = 0.76) based on the calculated 
Chandra ACIS-S point source sensitivity limit (4 $\times 10^{-15}$ ergs cm$^2$ s$^{-1}$ in 10$^4$ sec of exposure time and the redshifted [O III] flux from the spectral decomposition).
The spectral line ratio of log([O III]/H$\beta$) is 0.99 and places the southern nucleus as a Seyfert 2 spectral type.
The non-detection of H$\alpha$ broad-line and the X-ray in the Compton thin southern nucleus
suggest it could be a low-luminosity AGN (LLAGN; $L_{H\alpha} < 10^{42}$ erg s$^{-1}$).

The stellar velocity dispersions $\sigma_*$ of the southern and northern nuclei estimated from [O III] line width (Nelson \& Whittle 1996, Greene \& Ho 2005)
are 165 \kms and 117 \kms , respectively.
The black hole masses $M_{BH}$ of the southern and northern nuclei estimated from the $M_{BH}$ - $\sigma_*$ relation of
elliptical galaxies (Kormendy \& Ho 2013) are log($M_{BH}/M_\odot$)=8.11 and log($M_{BH}/M_\odot$)=7.47, respectively.
The nuclear separation between two black holes is 1.26 kpc.
If we assume they are moving in a circular orbit, then the orbital velocity from Kepler's law  is only 23 \kms.
If this is the case, then the observed velocity difference between two black holes will range from 0 to 46 \kms \ depending on the locations of the two SMBHs in the orbit and its inclination angle.
The expected observed velocity difference will be $\sim9$ \kms and can be calculated from the following expression:
\[ {bf 46}\ \rm{km\ s}^{-1} \times (\frac{2}{\pi})^2 \int_{0}^{\frac{\pi}{2}}\int_{0}^{\frac{\pi}{2}}\sin\theta\cos\phi\ d\phi d\theta, \]
where $\theta$ and $\phi$ are azimuthal angle and polar angle, respectively.
However, the velocity difference measured from [O III] lines in the spectra is 233 $\pm$ 30 \kms ($\Delta \lambda=3.9 \pm 0.5$ \AA).
If assuming two SMBHs orbiting each other, this velocity requires that the sum of two SMBH mass is larger than $10^9 M_\odot$. 
We argue that the offset AGN is not gravitationally bound with the black hole in the nucleus,
but subject to gravitational potential of the host galaxy (i.e. not a binary black hole, but a dual black hole system).

If a galaxy surface brightness profile follows S{\' e}rsic model, 
we can express a ratio of the enclosed luminosity with radius $r$ ($L(<r)$) and the total luminosity ($L(total)$) as follows (e.g., Ciotti 1991; Ciotti \& Bertin 1999; Graham \& Driver 2005; Yoon et al. 2011):
\[ L(<r)=L(total)\times {\frac{\gamma (2n, x)}{\Gamma (2n)}}, \]
where $\gamma$, $\Gamma$, and $n$ are incomplete gamma function, complete gamma function, and S{\' e}rsic index, respectively, 
and $x$ is defined as follows, using shape parameter $\kappa$ and effective radius $r_e$:
\[ x=\kappa (\frac{r}{r_e})^{\frac{1}{n}}. \]
The shape parameter $\kappa$ can be determined from the following formular (Ciotti \& Bertin 1999; MacArthur et al. 2003):
\[ \kappa \sim 2n - \frac{1}{3} + \frac{4}{405n} + \frac{46}{25515n^2} + \frac{131}{1148175n^3} - \frac{2194697}{30690717750n^4} + O(n^{-5}). \]
Then, if the stellar mass distribution follows the surface brightness distribution, the mass enclosed within the radius of $r$ can be calculated by the following expression:
\[ M(<r)=M(total)\times {\frac{\gamma (2n, x)}{\Gamma (2n)}}. \]
If the observed velocity difference is due to circular motion of the offset AGN with respect to the host galaxy's stellar mass that is enclosed within a 1.26 kpc radius, 
then this enclosed mass should be $\sim1.5\times10^{10} M_\odot$.
By using $\kappa$, $n$ and $r_e$ (1.26 kpc) from Galfit (\S 3.2.1), the ratio between the mass enclosed within 1.26 kpc and the total mass is 7.4\%, 
which implies that the total stellar mass of the host galaxy is $\sim 2.0 \times 10^{11} M_\odot$.
The SDSS $I$ band luminosity of the galaxy is $2.6\times10^{11} L_\odot$.
If we adopt $I$ band mass-to-light ratio of 1.7 (Shan et al. 2015), the estimated stellar mass based on $I$ band luminosity is consistent with the inferred host galaxy mass.


\subsection{Properties of the offset AGN (northern nucleus) and obscured AGN (host galaxy center)}

\subsubsection{Optical morphological properties}

In the previous paper (Kim et al. 2017), we have performed a two-dimensional galaxy fitting with GALFIT 3.0 (Peng et al. 2010)
and found the host galaxy is a bulge-dominated elliptical galaxy.
In the fitting, we have used a single point spread function (PSF) component that corresponds to the offset AGN.
Now, we have detected an optically hidden AGN in the center of the host galaxy and therefore 
have tried fitting the galaxy with two PSF components.
However, introducing an additional PSF component does not produce a good fitting result,
suggesting that the AGN in the host galaxy is deeply buried behind dust obscuration.
The nondetection of the second broad line component in the KECK spectra supports this result.
The host galaxy of the system can be well fitted with two galaxy components with two center positions: one centered at offset AGN and another centered at host galaxy center (top panel of Fig. 4).
A Sersic index of n=4 fits well to the southern galaxy that centers around the hidden AGN,
and a Sersic index of n=1 fits well to the northern galaxy that centers around the offset AGN.
Compared to the southern galaxy, which has a magnitude of m$_i$=16.82, the magnitude of the northern galaxy is m$_i$=19.07, indicating that the latter is significantly less luminous.
If we assume both galaxy components have the same mass to light ratio, then their mass ratio is about 8 (black hole mass ratio is about 4.4).
In addition, effective radius of the northern galaxy ($r_e$=0\farcs35) is 
significantly smaller compared to that of the southern nucleus ($r_e$=2\farcs5).
The large differences in mass and effective radius between the two galaxies 
and a lack of long tidal tails that typically observed in major mergers (i.e. the Antennae galaxies) suggest this system is a result of minor merger.

In the fitting, we have set centers of southern and northern galaxies as a free parameter (two center positions) for the best fitting result.
As a comparison, we have performed the Galfit with only one center position (i.e. the southern and northern galaxies have the same center position)
and find that the result is comparable ($\chi^2=1.60$ vs. $\chi^2=1.63$ for the fit with two center positions and one center position, respectively), 
but slightly worse than that of the two center positions (bottom panel of Fig. 4).
This result is expected due to the proximity of the two nuclei and the dominance of n=4 component in the fitting (i.e. coordinates of one center position falls near the center of the n=4 component).

\subsubsection{Radio Properties of the Hidden AGN}

The flux densities measured in the hidden AGN are 0.85 mJy, 0.63 mJy, and 0.24 mJy for 1.54 GHz, 10 GHz, and 33 GHz, respectively, assuming the VLBA is not resolving out any emission.
The derived spectral indicies $\alpha$ ($S_{\nu} \propto \nu^{-\alpha}$) are 0.16 and 0.81 for 1.54 GHz to 10 GHz and 10 GHz to 33 GHz, respectively.
As suggested by the high brightness temperature $T_b=7.2\times10^{7}$ K,
the radio emission at 1.54 GHz would be dominated by non-thermal synchrotron emission.
Thus, the flat spectral index at low frequencies can be explained
by Synchrotron self-absorption in an optically thick environment as observed in many LLAGNs (i.e. Nagar et al. 2000).
The steep spectral index of $\alpha=0.81$ between 10 GHz and 33 GHz is typical for Synchrotron emission (Gioia et al. 1982; Klein et al. 2018).
With decreasing frequency, the system is getting into a regime where the Synchrotron self-absorption is important and 
the spectral index becomes flat and turns over. 
The Synchrotron self-absorption that presumably occurs in the environment of large optical depth could explain
a hidden AGN in the host galaxy, which is not seen in optical but seen in radio, 
in contrast to the off-nucleus AGN which is bright in optical but does not have a radio counterpart.

The radio source in the southern nuclei was resolved with the VLBA and its deconvolved size is 3.4 mas $\times$ 2.4 mas corresponding to a physical scale of 17 pc $\times$ 12 pc.
The 1.54 GHz radio emission detected by the VLBA is originating from this small region. 
However, the flux density at 1.4 GHz (0.85 mJy) detected from the VLA Faint Images of the Radio Sky at Twenty centimetres 
(FIRST: Becker et al.  1995) survey is 1.61$\pm$0.15 mJy. 
This is twice as large as the value detected with the VLBA at 1.54 GHz.
The large synthesized beam size of FIRST (5\farcs 4) that captures almost all of the radio emission in the galaxy 
could explain the difference of the flux densities.
On the other hand, the intrinsic variability of the LLAGN could also explain the flux density difference (i.e. typical radio variability in LLAGN on a time scale of a few years is 20\% to 70\%: Falcke et al. 2001).
The far-infrared (FIR) luminosity of this galaxy estimated from the IRAS measurements is L$_{FIR}=10^{12.20}$L$_\odot$ 
and suggests the presence of significant star formation activity. 
Therefore, the additional radio emission of $\sim$0.75 mJy measured by FIRST could be coming from the star formation-related activity (i.e. synchrotron radiation from Type II supernovae).
The mean value of the $q_{24}$ parameter (log($S_{24\mu m}/S_{1.4 \rm{GHz}}$); Ibar et al. 2008) of the star-forming galaxies is 0.95.
The Spitzer/MIPS 24 $\mu m$ flux density for CXO J101527.2+625911 is 6.3 mJy.
If we assume that the excess 1.4 GHz emission measured by FIRST relative to the VLBA
(=0.75 mJy) is all due to star formation, then $q_{24}$ = 0.92; i.e., comparable to the $q_{24}$ of star-forming galaxies.
It could be possible that some fraction of the 24 $\mu m$ flux density could come from offset AGN. In such a case, the $q_{24}$ value could be smaller than measured.
 
\section{Summary}
We have carried out high resolution radio observations of CXO J101527.2+625911 
previously claimed as a rSMBH candidate with the VLA and the VLBA and found the followings:

$\bullet$
A single compact radio source was detected from the VLA and the VLBA observations near the center of the host galaxy.
The radio source is spatially resolved (17 pc $\times$ 12 pc) with the VLBA and its brightness temperature is $T_b=7.2\times10^{7}$ K, 
which clearly indicates that the radio emission is from AGN.
This combined with the Chandra-detected X-ray source associated with the northern offset AGN confirms that this system is a dual SMBH galaxy, not a rSMBH.

$\bullet$
Two narrow emission lines of [O III]$\lambda$5007\AA \ that were not seen in the SDSS spectrum were detected in the high resolution KECK spectra, 
supporting that this system is a dual SMBH galaxy.
The velocity offset between the two narrow lines of [O III]$\lambda$5007\AA \ is 233 $\pm$ 30 km s$^{-1}$.

$\bullet$
Morphological analysis of the HST image suggests this system has resulted from a minor merger, with a mass ratio of $\sim$8.
The nuclear separation between the two nuclei is 0\farcs26 $\pm$ 0\farcs01 (1.26 $\pm$ 0.05 kpc).


\noindent
\section{Acknowledgements}

\noindent
The authors thank the anonymous refree for comments and suggestions that greatly improved this paper.
We also thank T. Treu and J.-H. Woo for sharing their reduced Keck spectra, and K. Nyland for helping preliminary reduction of VLBA data.
This research has made use of the NASA/IPAC Extragalactic Database
(NED) which is operated by the Jet Propulsion Laboratory, California
Institute of Technology, under contract with the National Aeronautics and Space Administration. 
D.C.K., A.S.E., and S.S acknowledge support from the National Radio Astronomy Observatory (NRAO) and 
M.K. was supported by the National Research Foundation of Korea (NRF) grant funded by the Korea government (MSIT) (No. 2017R1C1B2002879). 
The National Radio Astronomy Observatory is a facility of the National Science Foundation operated under cooperative agreement by Associated Universities, Inc.

\begin{deluxetable}{lcccccccccc}
\tabletypesize{\small}
\tabletypesize{\scriptsize}
\tabletypesize{\tiny}
\tablewidth{0pt}
\tablecaption{Summary of VLBA and VLA Observations}
\tablehead{
\multicolumn{1}{c}{Tel.} &
\multicolumn{1}{c}{Date} &
\multicolumn{2}{c}{Position (J2000)} &
\multicolumn{1}{c}{Band} &
\multicolumn{1}{c}{On-source} &
\multicolumn{1}{c}{Beam} &
\multicolumn{1}{c}{PA} &
\multicolumn{2}{c}{Flux density} &
\multicolumn{1}{c}{RMS noise} \\
\multicolumn{1}{c}{} &
\multicolumn{1}{c}{} &
\multicolumn{1}{c}{R.A.} &
\multicolumn{1}{c}{Dec.} &
\multicolumn{1}{c}{} &
\multicolumn{1}{c}{hour} &
\multicolumn{1}{c}{\arcsec} &
\multicolumn{1}{c}{deg.} &
\multicolumn{1}{c}{peak} &
\multicolumn{1}{c}{total} &
\multicolumn{1}{c}{mJy/beam} \\
\multicolumn{1}{c}{(1)} &
\multicolumn{1}{c}{(2)} &
\multicolumn{1}{c}{(3)} &
\multicolumn{1}{c}{(4)} &
\multicolumn{1}{c}{(5)} &
\multicolumn{1}{c}{(6)} &
\multicolumn{1}{c}{(7)} &
\multicolumn{1}{c}{(8)} &
\multicolumn{1}{c}{(9)} &
\multicolumn{1}{c}{(10)} &
\multicolumn{1}{c}{(11)} 
}
\startdata
VLBA& 10/15/2017 & 10h15m27.267s & 62d59m11.350s & L (0.256) & 2.0 & 0.0125$\times$0.005    & 175.7 & 0.74$\pm 0.03$ & 0.85$\pm 0.05$ & 0.03 \\
VLA & 03/07/2018 & 10h15m27.268s & 62d59m11.348s & Ka (8.0)  & 0.6 &       0.06$\times$0.05 &  58.3 & 0.26$\pm 0.01$ & 0.24$\pm 0.01$ & 0.008 \\
VLA & 03/07/2018 & 10h15m27.270s & 62d59m11.351s &  X (4.0)  & 0.5 &       0.22$\times$0.17 &  55.1 & 0.61$\pm 0.03$ & 0.63$\pm 0.05$ & 0.005 \\
\enddata

(1) Telescope used. (2) Observing date. (3) Right Ascension. (4) Declination. (5) Radio band (Bandwidth in GHz unit): L: 18 cm, 1.54 GHz, Ka: 1 cm, 33.0 GHz, and X: 3.6 cm, 10.0 GHz. (6) On-source exposure time. (7) Synthesized beam size. (8) Position angle. (9) Peak (mJy/beam) and total (mJy) flux density. (10) RMS noise achieved.

\end{deluxetable}

\begin{figure}
\centerline{\includegraphics[scale=1.0]{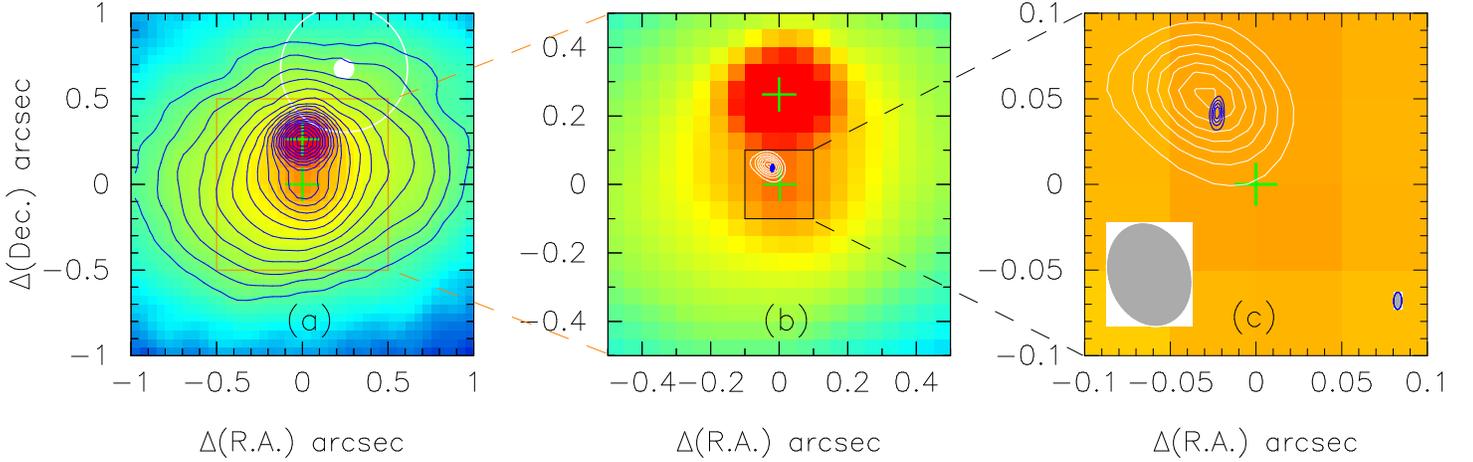}}
\caption{(a) The HST ACS I-band image of CXO J101527.2+625911,
(b) Zoom-in image of the boxed region in panel (a), and
(c) Zoom-in image of the boxed region in panel (b).
The white dot and circle in panel (a) represent the location of the Chandra X-ray source and its positional uncertainty, which is 0\farcs32.
The green plus signs represent two nuclei: a northern nucleus (offset AGN) and a southern nucleus (nucleus in the center of the host galaxy).
The white and blue contours in panel (b) represent VLA (33.0 GHz) and VLBA (1.54 GHz) continuum emission, respectively.
The inset box in panel (b) is a zoomed-in view of the nuclear region.
The synthesized beam sizes are plotted on the bottom-left (JVLA) and bottom-right (VLBA) in panel (c).
Contour levels are spaced in log 0.15 units apart.
North is up and east is to the left.
}

\end{figure}


\begin{figure}[!hb]
\centerline{\includegraphics[scale=0.6]{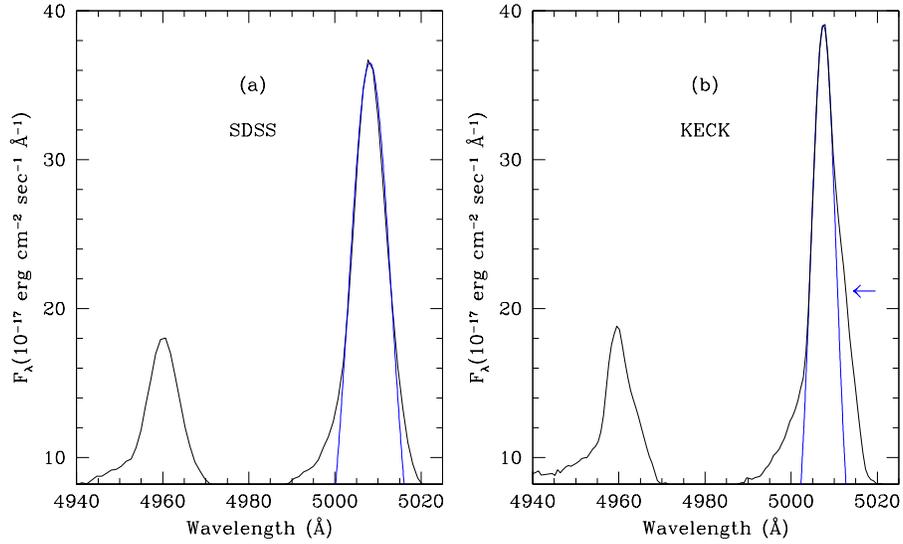}}
\caption{[O III]$\lambda$5007 line in SDSS (a) and KECK (b) spectra fitted with a single Gaussian profile.
An additional component is shown in the red part of the [O III]$\lambda$5007 line in the KECK spectra, suggesting the existence of a second ionizing source.
}

\end{figure}

\begin{figure}[!ht]
\centerline{\includegraphics[scale=1.0]{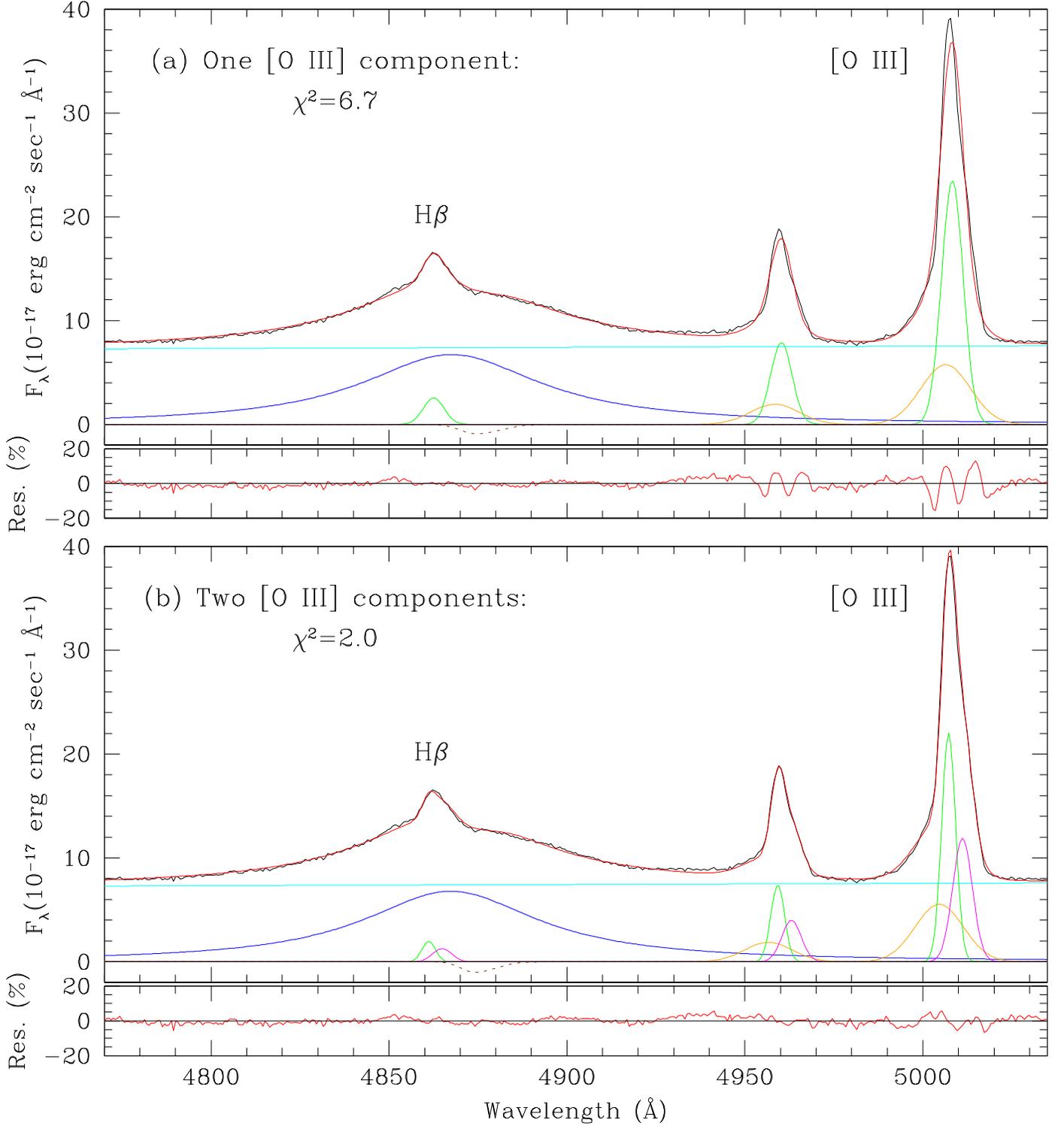}}
\caption{ Result of spectral decompositions of the H${\beta}$ + [O III] region using a single [O III] component (a) and a double [O III] component (b).
Black, cyan, blue, and red lines represent
the data, power-law continuum model fit, broad emission line model fit, and the combined model fit, respectively.
Green and magenta represent fit to the narrow emission lines and orange represents [O III] blue asymmetry component.
The model fit with two [O III] components yields better result.
}

\end{figure}

\begin{figure}
\centerline{\includegraphics[scale=0.82]{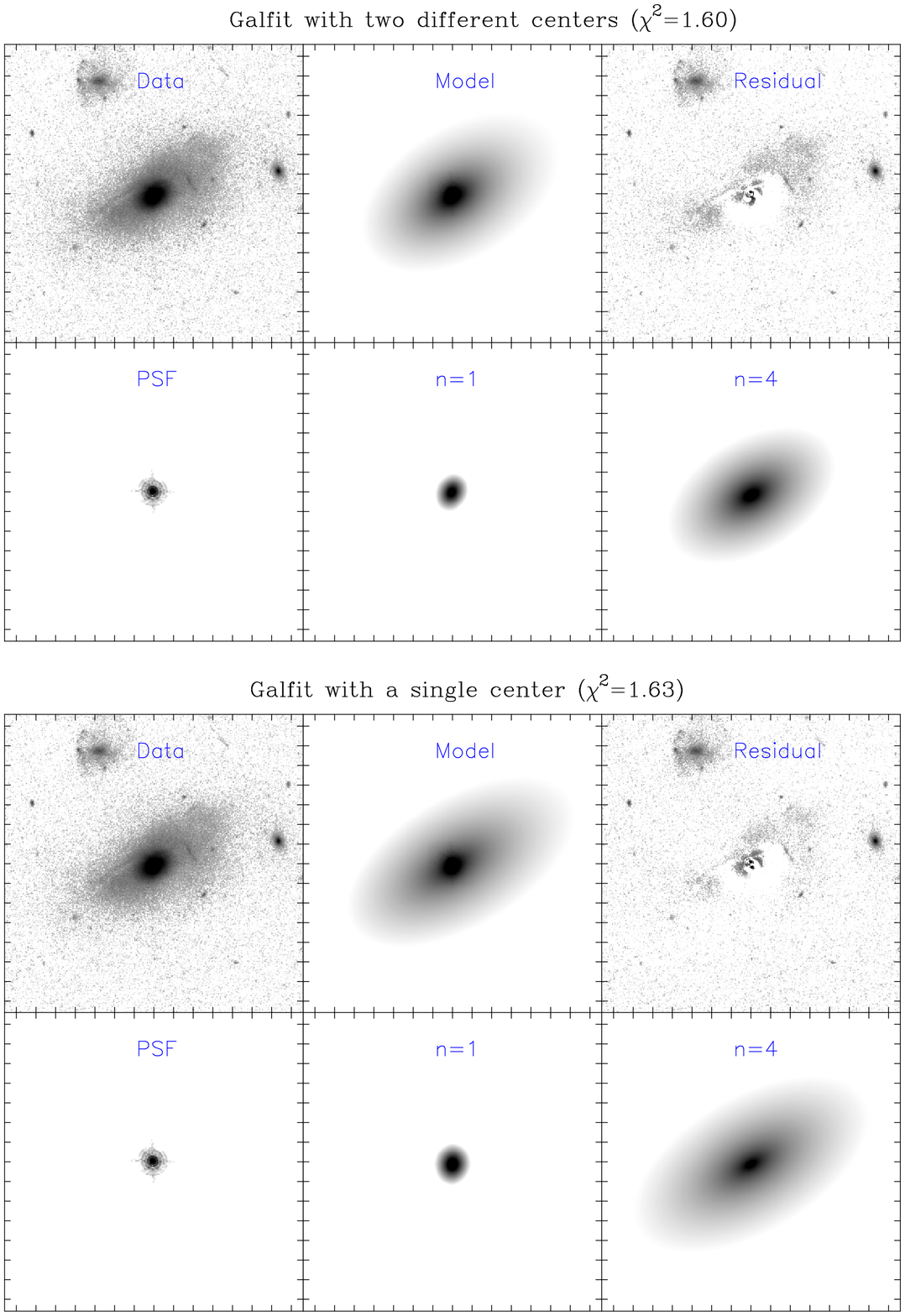}}
\caption{Results of Galfit with two centers (top panel) and with a single center (bottom panel). There is no significant difference between the two fits.
}

\end{figure}

\clearpage

\end{document}